\documentclass[aps,prl,groupedaddress,notitlepage]{revtex4-1}
\usepackage{fullpage}
\usepackage{graphicx} 
\usepackage{amsmath}
\usepackage{amsfonts}
\usepackage{amssymb}
\usepackage{hyperref}
\usepackage{float}
\usepackage{mdframed}
\usepackage{color}
\definecolor{explanationcolor}{RGB}{240,240,255}
\definecolor{examplecolor}{RGB}{240,240,240}
\definecolor{exercisecolor}{RGB}{250,240,240}
\definecolor{problemcolor}{RGB}{225,255,240}
\definecolor{challengecolor}{RGB}{255,240,225}
\definecolor{relevlitcolor}{RGB}{250,250,250}
\definecolor{hyperlinkcolor}{RGB}{0,0,0}
\definecolor{hypercitecolor}{RGB}{0,180,90}
\usepackage{hyperref}

\newenvironment{explanationbox}{\begin{mdframed}[backgroundcolor=explanationcolor,linewidth=1pt]}{\end{mdframed}}
\newenvironment{examplebox}{\begin{mdframed}[backgroundcolor=examplecolor,linewidth=0pt]}{\end{mdframed}}
\newenvironment{exercisebox}{\begin{mdframed}[backgroundcolor=exercisecolor,linewidth=0pt]}{\end{mdframed}}

\newcounter{explanation}
\def\theexplanation{\arabic{explanation}}

\newcounter{example}
\def\theexample{\arabic{example}}

\newcounter{exercise}
\def\theexercise{\arabic{exercise}}
\newenvironment{exercise}[1][]{\begin{exercisebox}\refstepcounter{exercise}\par\medskip
   \noindent \textbf{Exercise~\theexercise:} #1}{\medskip\end{exercisebox}}

\newcounter{problem}
\def\theproblem{\arabic{problem}}
\newenvironment{problem}[1][]{\refstepcounter{problem}\par\medskip
   \noindent \textbf{Problem~\theproblem:} #1}{\medskip}
    
\begin{document}

\author{Felix Ritort}
\email{ritort@ub.edu,fritort@gmail.com}
\affiliation{Small Biosystems Lab, Condensed Matter Physics Department, University of Barcelona, C/Mart\'{\i} i Franqu\'es 1, 08028 Barcelona, Spain; and\\
Institut de Nanoci\`encia i Nanotecnologia (IN2UB), Universitat de Barcelona, 08028 Barcelona, Spain}


\title{Nonequilibrium work relations for energy and information}

\maketitle

{\it I review basic concepts in the nonequilibrium physics of small systems, emphasizing single molecule experiments and how they contribute to expanding our current understanding of energy and information. }
\section*{Introduction}
\label{FTs:intro}
\begin{figure}[b!]
	\begin{center}
		\includegraphics[height=6cm]{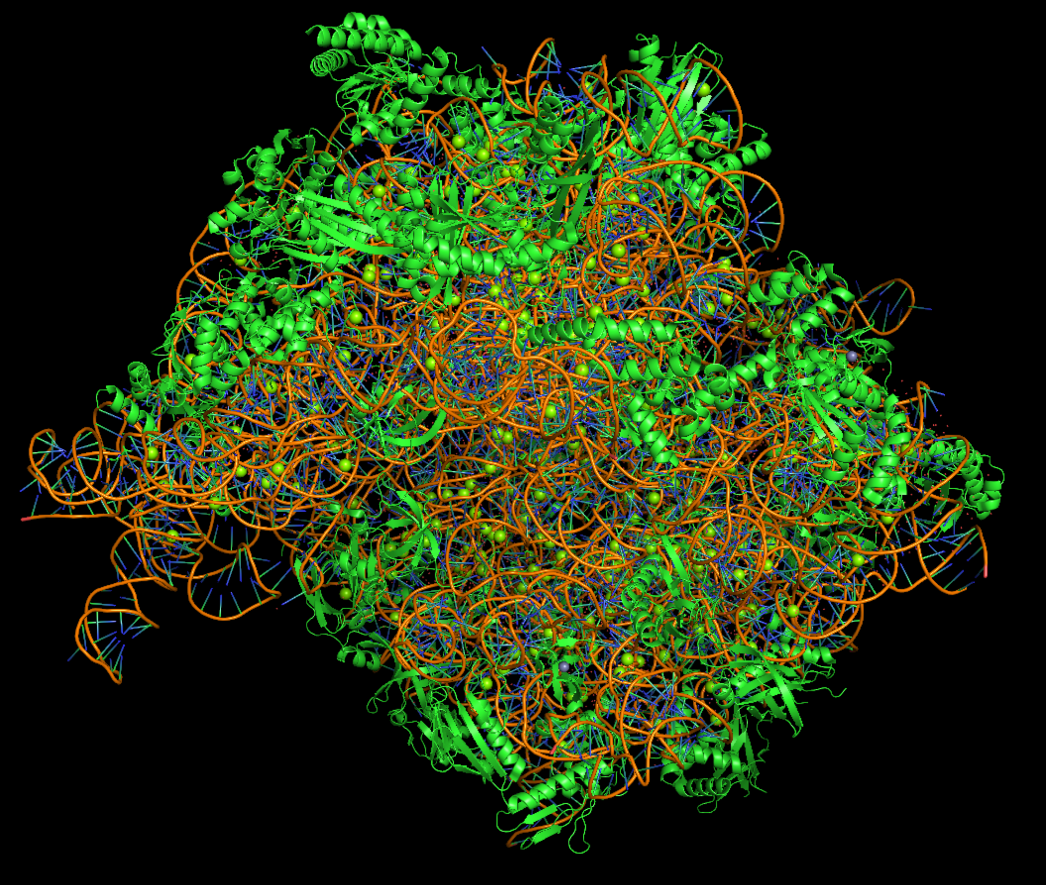}		
	        \caption{
	        {\bf Structure of the ribosomal unit 70S from {\em E. coli} obtained with cryo-electron microscopy at 2$\AA$ resolution.} One of the best characterized and highly conserved across prokaryotes of the biological platform in charge of synthesizing proteins from the messenger RNA. The ribosomal machine is a large nucleoprotein complex made of RNA and protein that weighs about 2.3-4 Mega Daltons. One nucleotide and amino acid being about 100 Daltons, means that this marvelous of nature contains tens of thousands of nucleotides and aminoacids. Figure taken from Ref.\cite{watson2020structure}    }\label{fig:FTs:ribosome}
	\end{center}
\end{figure}


Non-equilibrium pervades nature. Most natural systems are out of equilibrium, from a waterfall to a star, from a microbe to a human being. A system is out of equilibrium when non-zero currents of any conserved quantity (e.g., mass, charge, momentum, energy) are present. Let us think of a block sitting on the floor. If we pull on it, the frictional forces heat the block at the contact area, generating a heat flow toward the floor. The exerted mechanical power is irreversibly transformed into heat $\rlap{\textrm{d}}{\bar{\phantom{w}}}Q$ during a time $dt$, producing entropy at a rate $\sigma=\frac{1}{T}\rlap{\textrm{d}}{\bar{\phantom{w}}}Q/dt\ge 0$, where $T$ is the temperature of the environment. Nonequilibrium systems have (positive) entropy production rate $\sigma$, whereas in equilibrium $\sigma=0$. If the opposite is observed, i.e., heat spontaneously flows from the floor to the block, the entropy production is negative during $dt$, $\sigma<$0. Such events are rare. If observed, they are evidence that entropy must have been produced elsewhere to compensate for such a decrease. According to the second law of thermodynamics, the global balance of entropy production in the universe is always positive. 

Moreover, if the pulled body is small enough, for example, a charged dust particle driven by an electric field and subject to friction, or the time interval $dt$ is too short, the entropy production rate is sometimes negative, meaning that heat spontaneously flows from the surroundings to the body. Upon decreasing size and time, such negative events become more frequent and entropy production fluctuations larger. Ultimately, irreversibility fades away in the limit of exceedingly short times below the molecular collision timescale, and motion appears reversible. 

These behaviors emerge from the so-called Brownian motion \cite{zwanzig2001nonequilibrium}. In 1827, Robert Brown, a botanist well known for his detailed descriptions of the cell and contributions to plants' taxonomy, made an important discovery during his pollination studies. While examining the motion of the grains of pollen suspended in water through the microscope, he observed the motion of grains to be erratic, as if they were alive. Instead, Brown had observed the effect of the random collisions of the water molecules against the grains of pollen: kicked from all directions, the pollen grains jiggled erratically in the water solution.  Statistical physics builds on the atomistic nature of matter and the motion-like origin of heat and work. The equipartition law sets $k_BT$ ($k_B$ being Boltzmann's constant) as the energy scale where molecular motion and thermal fluctuations are observed \cite{chandler1987introduction}. The branch of physics that studies nonequilibrium processes in systems with a few degrees of freedom is called Nonequilibrium Thermodynamics of Small Systems \cite{bustamante2005nonequilibrium} and Stochastic Thermodynamics \cite{seifert2012stochastic,ciliberto2017experiments}. Prominent examples are colloidal particles captured in optical traps, biomolecules pulled under mechanical forces, and single-electron transistors. With the development of high-temporal resolution cameras, photomultipliers, and photodetectors for light detection, our ability to measure energy fluctuations has expanded. This paper reviews some of the concepts and applications in this exciting field. To complement the theory, I propose a few selected exercises and problems.

\section{Small systems and single molecule experiments}\label{FTs:sme}
Single-molecule biophysics has become a fabulous playground to investigate the nonequilibrium processes in small systems in physics and biology \cite{ritort2008nonequilibrium,dieterich2015single}.  Several techniques permit us to manipulate individual biological molecules one at a time, such as nucleic acids and proteins \cite{ritort2006single,neuman2007single,moffitt2008recent,ritchie2015probing,miller2017single,bustamante2021optical}. Living matter is soft and stabilized by weak molecular forces of entropic origin, such as electrostatic and hydrophobic, of typical energies on the order of $k_BT$. Since these energies are comparable to the thermal forces, bio-matter is suitable for observing large fluctuations and negative entropy events. Laser optical tweezers for single-particle and single-molecule manipulation have become a crucial technique in the field \cite{gieseler2021optical}. Invented by Arthur Ashkin in Bell Labs in 1970 and awarded the Nobel Prize in Physics in 2018, optical tweezers have revolutionized physics, chemistry, and biology research.

\begin{figure}[b!]
	\begin{center}
		\includegraphics[height=8cm]{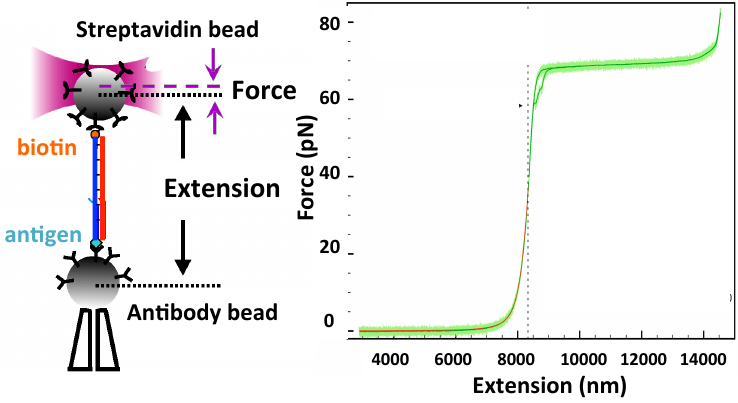}		
	        \caption{
	        {\bf Force-extension curve (FEC) for DNA.} Experimental setup in a DNA pulling experiment. (Right) FEC was measured by pulling a 24kb DNA molecule (half $\lambda$-DNA) at standard conditions ($T=298$K and 1M NaCl Tris buffer). Three force regimes are identified. At forces below 10pN the DNA random coil is extended against thermal fluctuations (entropic regime). At forces between 10pN and 60pN the DNA is stretched above its contour length (8.3$\mu$m) (enthalpic regime). Above 65pN, DNA is overextended by $\sim 70\%$ of its contour length.}
	
	   \label{fig:FTs:DNAstretching}
	\end{center}
\end{figure}

Optical traps for single-molecule manipulation are produced by focusing an infrared beam inside a fluidics chamber, optically trapping a micrometer-sized bead, and measuring the force from the deflected light using position-sensitive detectors. In particular, counter-propagating traps measure forces by directly measuring the change in light momentum \cite{smith20037}.  {\it “Take a single DNA molecule and pull from its extremities while recording the force-extension curve until it gets fully straightened”}.  This thought experiment, a dream a few decades ago, has become standard in many research institutes worldwide.  Labeling the ends of a DNA molecule with specific chemical groups (biotin, avidin, digoxigenin) makes it possible to tether a single DNA between two beads \cite{smith1996overstretching}. Moving one bead relative to the other and using it as a force sensor makes it possible to measure a single biopolymer's force-extension curve (FEC), from DNA to RNA and proteins. In single-trap setups, one bead is immobilized in a pipette by air suction; the other is captured in an optical trap and measures the force exerted on the molecule as a function of the trap position. Figure \ref{fig:FTs:DNAstretching} shows the FEC obtained by pulling a 24kb fragment of the DNA of bacteriophage lambda, a virus that infects bacteria, and a model in molecular biology for decades. The measured elastic response is described by semiflexible polymer models such as the worm-like chain (WLC) model, a simplified version of the more general elastic rod model \cite{nelson2008biological}. The key ingredient of the WLC is the bending stiffness $A$ or the energy cost to locally bend the polymer, which is proportional to the persistence length $P$, $A=k_BTP$. The energy of a configuration in the WLC is given by, 
\begin{equation}
    \label{eq:sme:WLCenergy}
 E=\frac{A}{2}\int_0^Lds\Bigl(\frac{\partial \hat{t}}{\partial s}\Bigr)^2   
\end{equation}
where $L$ is the contour length, $s$ is the arc coordinate and $\hat{t}(s)$ is the tangent unit vector field, see the Exercise \ref{ex:FTs:WLC}.

\begin{exercise}\label{ex:FTs:WLC}
{\bf Persistence length in the WLC model.}
Let us consider Eq.\eqref{eq:sme:WLCenergy}. 

{\bf a.} Compute analytically the partition function ${\cal Z}=\int {\cal D}\hat{t}(s)\exp(-\beta E)$ either by Fourier transforming the tangent vector field or by discretizing the polymer chain and mapping the resulting discrete model to a classical 1D Heisenberg model without external field. 

{\bf b.} Compute the tangent unit vector correlation function in the thermodynamic limit $L\to\infty$, $C(s)=\langle \hat{t}(0)\hat{t}(s)\rangle$ demonstrating that $C(s)=\exp(-s/P)$. This shows that the persistence length is the distance along the contour length over which directionality is lost. 

{\bf c.} Finally, demonstrate that the polymer end-to-end distance, defined as $\Vec{R}=\int_0^Lds\hat{t}(s)$ satisfies, $\langle\Vec{R} \rangle=0$ and $\langle\Vec{R}^2 \rangle=2LP$.
{\it [Hint: Show first that $\langle\Vec{R}^2 \rangle=2L\int_0^LdsC(s)$.]}.

\end{exercise}
Persistence length measurements in biopolymers are often done with electrophoresis, light diffraction, AFM, and fluorescence imaging. Accurate estimations of $P$ are obtained by measuring the FEC in single-molecule pulling experiments and fitting it to the WLC model. Marko and Siggia derived a formula interpolating low and high forces two decades ago \cite{marko1995stretching} (Problem \ref{prob:FTs:WLC}). It is given by,
\begin{equation}
    \label{eq:sme:WLCMarkoSiggia}
 f(x)=\frac{k_BT}{P}\Bigl( \frac{1}{4(1-x/L)^2}-\frac{1}{4}+\frac{x}{L}\Bigr)\,\, .
\end{equation}
It is common to include the extensibility of the polymer's contour length by introducing a stretching term in Eq.\eqref{eq:sme:WLCenergy}. This leads to the extensible-WLC, described by Eq.\eqref{eq:sme:WLCMarkoSiggia} with the rescaling $L\to L(1+f/Y)$ where $Y$ is the Young modulus \cite{bustamante1994entropic}. The resulting implicit equation can be inverted (to obtain $x(f)$) by mapping it to a third-degree equation \cite{severino2019efficient}. Figure \ref{fig:FTs:DNAstretching} shows a WLC fit (red line) to the experimental data (green line) between 0 and 40pN that reproduces the data pretty well. Values at standard conditions (298K,1M NaCl) are $P=50$nm, $Y=1000$pN for double-stranded DNA (dsDNA), $P=0.75$nm for single-stranded DNA (ssDNA) \cite{bosco2014elastic,jacobson2017single} and single-stranded RNA (ssRNA) \cite{bizarro2012non},  $P=0.6$nm for polypeptide chains, etc. The WLC model does not consider self-exclusion effects, which are essential at low salts \cite{saleh2009nonlinear,viader2021cooperativity}, and other models are needed, such as the thick-chain model \cite{toan2005inferring,toan2006inferring}). Elastic measurements have also been performed at different temperatures using a temperature-controlled optical trap \cite{de2015temperature}. While $P$ decreases linearly with $T$ for dsDNA, it does increase linearly with $T$ for ssDNA \cite{rico2022temperature}
\section{Nonequilibrium force spectroscopy}\label{FTs:fs}
\begin{figure}[b!]
	\begin{center}
		\includegraphics[height=8cm]{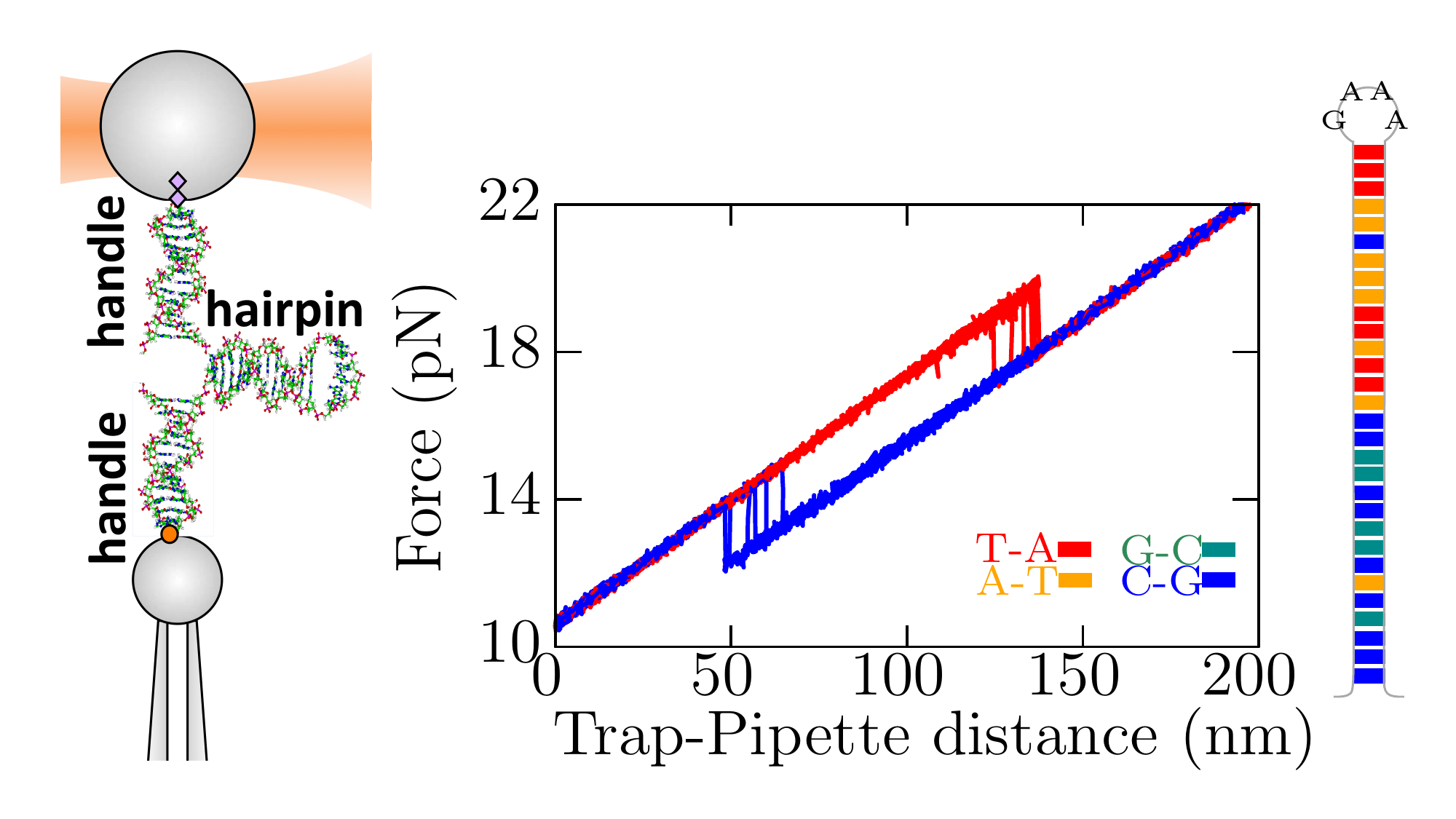}
	        \caption{
	        {\bf Mechanical unzipping experiments.} (Left) Experimental setup showing the optical trap, the flanking DNA handles, and the DNA hairpin (figure not to scale). (Right) Force-distance curves measured in the mechanical unfolding and folding of DNA hairpins. The force rips correspond to unfolding (red) and folding (blue) transitions.	    }\label{fig:FTs:DNAunzipping}
	\end{center}
\end{figure}

Besides measuring polymer elasticity, single-molecule experiments provide a tool to investigate molecular folding of nucleic acids and proteins, DNA-protein and DNA-peptide interactions (e.g., intercalation, condensation, aggregation phenomena), protein-protein interactions (e.g., ligand-receptor binding), molecular motors (e.g., cellular transport, DNA-RNA polymerases, ATPases and proton pumps, viral packaging motors, topoisomerases, helicases, etc.). For example, folding thermodynamics and kinetics are investigated by unzipping molecular structures: molecular ends are pulled apart to break the bonds that hold the native structure \cite{essevaz1997mechanical,rief1999sequence}. Unzipping experiments on DNA hairpins of a few kb have permitted determining the hybridization energies of DNA \cite{huguet2010single,huguet2017derivation}, RNA \cite{rissone2022stem,rissone2022nucleic} and the binding sites of ligands \cite{manosas2017single} at single base pair resolution with 0.1kcal/mol accuracy. Mechanical unzipping has provided a single-molecule test of the functional renormalization group approach for disordered elastic systems \cite{huguet2009statistical,ter2023experimental}. In Figure \ref{fig:FTs:DNAunzipping} we show a typical unzipping experiment of a 20bp-DNA hairpin (a stem-loop structure made of two complementary DNA strands ending in a loop). A molecular construct of the hairpin linked to dsDNA handles is tethered between two beads for unzipping \cite{forns2011improving}. Moving the optical trap relative to the pipette, we measure force and work that give valuable information about the folding process, such as free energy, kinetic rates, and folding pathways \cite{manosas2009dynamic,mossa2009dynamic}. 
To characterize mechanical unfolding and folding stability, measuring the rupture force distributions where the molecule unfolds and refolds for the first time is customary. These distributions give information about the folding free-energy landscape of the molecule, such as the height of the kinetic barrier $B$, the folding free energy at zero force $\Delta G_0$, and the distances between the folded state and the transition state, $x^{\dagger}$, and the unfolded state and the transition state, $x^{\ast}$.  A successful model that fits the experimental data is the Bell-Evans model with force-dependent kinetic rates,
\begin{eqnarray}
    k_{F\to U}(f)=k_0\exp\bigl(\frac{fx^{\dagger}}{k_BT} \bigr)\label{eq:fs:BellEvans1}\\
    k_{U\to F}(f)=k_0\exp\bigl(\frac{\Delta G_0-fx^{\ast}}{k_BT} \bigr)\label{eq:fs:BellEvans2}
\end{eqnarray}
where $k_0$ is the unfolding rate at zero force. These rates fulfill detailed balance, $k_{F\to U}(f)/k_{U\to F}(f)=\exp[(fx_m-\Delta G_0)/k_BT]$ with $x_m=x^{\dagger}+x^{\ast}$.

A first-order Markov process yields the first rupture force distributions and the survival probabilities of the folded and unfolded states during unfolding and refolding, respectively. 
Particularly useful are the mean and variance of the rupture forces that can be derived within a Gaussian approximation, see the exercise \ref{ex:FTs:BellEvans}. By detecting cooperative changes in force and extension, force spectroscopy has become a powerful approach to identifying intermediate states and pathways in molecular folding \cite{alemany2017force,rico2022force}.
\begin{exercise}\label{ex:FTs:BellEvans}
{\bf Analytical rupture forces in unzipping experiments.} Let us consider a two-state molecule either folded or unfolded under an applied force. Unfolding and folding kinetic rates are given by Eq.\eqref{eq:fs:BellEvans1},\eqref{eq:fs:BellEvans2}. Initially, the molecule is folded at $f=f_{min}$, and the force is ramped at the pulling rate $r=\dot{f}$.  

{\bf a.} Write the master equation for the survival probability of the folded state during unfolding and derive the first rupture force distribution $\rho_{F\to U}(f)$,
\begin{equation}
    \rho_{F\to U}(f)=\frac{k_0}{r}\exp\Bigl(\beta x^{\dagger}f+\frac{k_0}{\beta r x^{\dagger}} \bigl(\exp(\beta x^{\dagger}f_{min})- \exp(\beta x^{\dagger}f)\bigr)\Bigr)
    \label{eq:fs:rhoFU}
\end{equation}
{\it [Hint: Express the master equation in terms of force rather than time]}.

{\bf b.} Use symmetry (F$\to$U) arguments to derive the corresponding expression for folding $\rho_{U\to F}(f)$, starting from a maximum force, $f_{max}$, where the molecule is unfolded with probability one. Plot the two functions, $\rho_{F\to U}(f),\rho_{U\to F}(f)$. {\it [Suggestion: For the plots, take $f_{min}=-\infty$ for $\rho_{F\to U}(f)$ and $f_{max}=\infty$ for $\rho_{U\to F}(f)$.]}. 

{\bf c.} Starting from Eq.\eqref{eq:fs:rhoFU}, find the most probable rupture force $f^{\ast}$ and the standard deviation $\sigma_f$ of the rupture force distribution using a Gaussian approximation, i.e. expanding Eq.\eqref{eq:fs:rhoFU} to second order around its maximum. Demonstrate that, 
\begin{eqnarray}
f^{\ast}=\frac{k_BT}{x^{\dagger}}\log\Bigl(\frac{x^{\dagger}r}{k_0 k_BT}\Bigr)\\
\sigma_f=\log \Bigl( \frac{3+\sqrt{5}}{2}\Bigr)\frac{k_BT}{x^{\dagger}}
\end{eqnarray}
{\bf d.} Estimate the values of $f^{\ast},\sigma_f$ for the DNA hairpin of 20bp shown in Figure \ref{fig:FTs:DNAunzipping} and check that $f^{\ast}$ and $\sigma_f$ qualitatively agree with the magnitude of the rupture forces and dispersion that you see in the figure (take $T=298$K, $x^{\dagger}=9$nm, $k_0= 10^{-16}$s$^{-1}$, $r=$20pN/s)
\end{exercise}

\section{Fluctuation theorems}
\label{FTs:fts}
Fluctuation theorems (hereafter referred to as FTs) quantify the probability of observing the rare negative entropy production events relative to the most probable positive ones. In their simplest form, they obey a simple mathematical relation \cite{evans2002fluctuation},
\begin{equation}
    \frac{P(S_t)}{P(-S_t)}=\exp\bigl( {\frac{S_t}{k_B}}\bigr )
\label{eq:fts:ft}
\end{equation}                      
where $S_t$ is the entropy produced during time $t$.  Equation \eqref{eq:fts:ft} shows that trajectories with $S_t<0$ are exponentially suppressed relative to those with $S_t>0$. Originally proven by Giovanni Gallavotti and Eddy Cohen for nonequilibrium steady states in the large $t$ limit \cite{gallavotti1995dynamical}, Eq.\eqref{eq:fts:ft} is also valid for transient states at any $t$ \cite{seifert2012stochastic}. Transient states are generated by driving out of equilibrium a system initially in equilibrium at $t=0$. Dynamics is modeled by the action of time-dependent external forces parametrized by $\lambda(s)$ with $0\le s\le t$. For a given process (forward, F), it is customary to define its time-reversed (R). In the reverse process, the system starts in equilibrium at $\lambda(t)$, and the control parameter evolves backward in time relative to the forward process, $\lambda_R(s)=\lambda(t-s)$. The transient work-FT takes the form,
\begin{equation}
    \frac{P_F(W_t)}{P_R(-W_t)}=\exp\bigl(\frac{W_t-\Delta G}{k_BT} \bigr)\,\,.
\label{eq:fts:ftcrooks}
\end{equation}
Here, $W_t$ equals the work exerted by the external driving forces during time $t$, and $\Delta G$ is the system's free energy difference between equilibrium states at $\lambda(0)$ and $\lambda(t)$. $\Delta G$ also equals the reversible work between the initial and final states. Equation \eqref{eq:fts:ftcrooks} is known as Crooks Fluctuation Theorem for work \cite{crooks1999entropy}. Equation \eqref{eq:fts:ftcrooks} can be related to \eqref{eq:fts:ft} taking $S_t=W_{\rm d}(t)/T$ with $W_{\rm d}=W-\Delta G$ the difference between the actual work and the reversible work, also called dissipated work. 
Note the difference between \eqref{eq:fts:ftcrooks} and \eqref{eq:fts:ft}; in the latter case, there is no distinction between forward and reverse. Another difference between \eqref{eq:fts:ftcrooks} and \eqref{eq:fts:ft} is the role of the so-called boundary terms. While \eqref{eq:fts:ftcrooks} is exact for all times, the more general \eqref{eq:fts:ft} is only valid in the large time limit. For nonequilibrium steady-states, $S_t$ grows linearly with $t$ on average, defining the entropy production rate $\sigma=\langle S_t\rangle/t$ where $\langle ..\rangle$ stands for an average over trajectories. Equation \eqref{eq:fts:ft} takes the form, 
\begin{equation}
\sigma=k_B\lim_{t\to\infty}\frac{1}{t}\log\Bigl(\frac{P(S_t)}{P(-S_t)}\Bigr)
\label{eq:fts:sigma}
\end{equation}
with finite $t$ corrections to $\sigma$ on the order of $1/t$.  To understand the origin of such corrections, we observe that heat $Q_t$ and work $W_t$ are related by energy conservation, $Q_t+W_t= \Delta U_t$, with the internal energy difference, $\Delta U_t=U_t-U_0$, a boundary term between the initial and final time. While $Q_t,W_t$ grow linearly with $t$ on average, $\Delta U_t$ does not, the difference between $Q_t$ and $W_t$ remains finite in the large $t$ limit, and of magnitude $1/t$ relative to either $Q_t$ or $W_t$. Therefore, while $S_t$ in Eq.\eqref{eq:fts:ft} stands for the heat produced $Q_t$, the equivalent quantity in Eq.\eqref{eq:fts:ftcrooks} stands for the exerted work $W_t$. The difference between both quantities remains a finite boundary term $\Delta U_t$, mostly contributing to the tails of the distributions. These tails can be described in the framework of the large deviations theory. We are not going to talk about them here. More about the role of the tails can be found in problem \ref{prob:FTs:steadystate}. 

 Equations \eqref{eq:fts:ft},\eqref{eq:fts:ftcrooks} imply two results. First, as we said, trajectories with negative $S_t$ are exponentially suppressed relative to the positive $S_t$ ones. The suppression of such trajectories exponentially grows with time and size due to the extensivity property of $S_t$ with time and size. Second, a corollary of Eq.\eqref{eq:fts:ftcrooks} is the Jarzynski equality \cite{jarzynski1997nonequilibrium}. If rewritten as $\exp(-W_t/k_BT)P_F(W_t)=P_R(-W_t)\exp(-\Delta G/k_BT)$ and integrating over $W_t$ we get $\langle\exp(-W_t/k_BT)\rangle=\exp(-\Delta G/k_BT)$ or $\langle\exp(-W_{\rm d}(t)/k_BT)\rangle=1$ where $\langle..\rangle$ denotes an average over many experimental realizations of the forward process. From Jensen's inequality or the convexity of the exponential function, $\langle\exp(x)\rangle\ge \exp(\langle x\rangle$, being $x$ a random variable,  it follows that $\langle W_{\rm d}(t)\rangle\ge 0$, in agreement with the second's law of thermodynamics. Similarly, one can prove that $\sigma\ge 0$ from Eq.\eqref{eq:fts:sigma}. The beauty of Eqs.\eqref{eq:fts:ft},\eqref{eq:fts:ftcrooks} lies in the fact that the second law is derived from a mathematical equality, rather than the less informative Clausius inequality.

\begin{figure}[b!]
	\begin{center}
		\includegraphics[height=5cm]{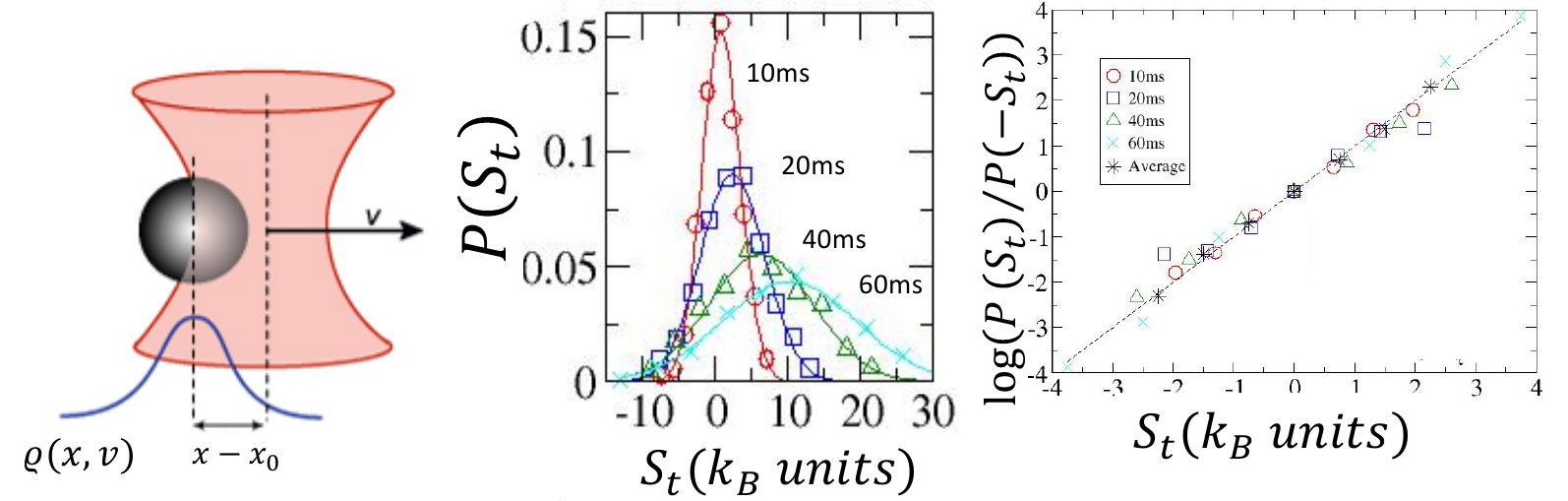}
	        \caption{
	        {\bf Bead in a trap dragged through water.} (Left) Experimental setup showing a bead captured in an optical trap moving at a speed $v\sim 1\mu$m/s. (Middle) Entropy production distributions are measured at different times. (Right) Experimental test of Eq.\eqref{eq:fts:ftcrooks}.	    }\label{fig:FTs:beadtrap}
	\end{center}
\end{figure}

The first experimental test of Eqs.\eqref{eq:fts:ft},\eqref{eq:fts:ftcrooks} was made in 2002 by Denis Evans and collaborators \cite{wang2002experimental}. They measured the entropy production in a microsphere optically trapped and dragged through water. Initially equilibrated in a resting optical trap, the bead is driven out of equilibrium by instantaneously moving the optical trap at a constant velocity $v$. The work exerted on the bead up to time $t$ is given by, $W_t=W_{\rm d}(t)=v\int_0^tf(s)ds$ where $\Delta G=0$ because the bead-trap potential is translationally invariant. Note that the reverse process requires reversing the optical trap's motion. For spatially symmetric traps, $V(x)=V(-x)$ and $P_F=P_R$. Evans and collaborators measured the entropy production distribution from the particle's trajectories finding trajectories with $S_t$ negative as predicted by Eq.\eqref{eq:fts:ftcrooks}. Some experimental results are shown in Figure \ref{fig:FTs:beadtrap}. An analytical proof of Eq.\eqref{eq:fts:ftcrooks} can be obtained from the Langevin equation describing the overdamped motion of a bead captured in a harmonic potential moving in the lab frame at speed $v$, 
\begin{equation}
    \gamma \dot{x}(t)=-k_b(x(t)-x_0(t))+\eta(t)
\label{eq:fts:lang}
\end{equation}
where $\eta$ is a white noise of zero mean and correlation $<\eta(t)\eta(s)>=2k_BT\gamma\delta(t-s)$, $k_b$ is the trap stiffness and $x_0(t)=vt$ is the center trap position. Similar results have been obtained for non-Markovian systems with memory effects \cite{di2020explicit}. Equation \eqref{eq:fts:lang} is linear and can be readily solved for an arbitrary initial condition finding $P(S_t)$; check the following exercise \ref{ex:FTs:bead}.

\begin{exercise}\label{ex:FTs:bead}
{\bf Entropy production of an optically trapped bead dragged through the water.} Let us consider a micro-sphere of radius $R$ immersed in water and optically trapped in the focus of a laser (typically in the infrared domain to avoid light absorption by water). The trapping potential can be considered as quadratic, $V(x)=(1/2)k_b(x-x_0)^2$ where $x_0$ is the trap center and $k_b$ is the trap stiffness. From Eq.\eqref{eq:fts:lang} do the following exercises,

{\bf a.} Demonstrate that in equilibrium ($x_0$ fixed), the autocorrelation function of the force, $C(t,s)=\langle f(t)f(s)\rangle$ with $f(t)=-k_b(x(t)-x_0)$ is given by,
\begin{equation}
    C(t,s)=C(t-s)=k_BT k_b \exp\bigl(-\omega_c (t-s)\bigr) 
    \label{eq:fts:cts}
\end{equation}
where $\omega_c=k_b/\gamma$ is the corner frequency of the bead that sets the relaxation timescale of the bead in the trap $\tau_r=1/\omega_c=\gamma/k_b$.

{\bf b.} Calculate the average and variance of the entropy production, $\langle S_t\rangle,\sigma_{S_t}^2=\langle S_t^2\rangle-\langle S_t\rangle^2$, and show that both grow linearly with time. Check that they fulfill the fluctuation-dissipation relation at any $t$,
\begin{equation}
\frac{\sigma_{S_t}^2}{2k_B\langle S_t\rangle}=1
    \label{eq:fts:fdr}
\end{equation}
{\it [Note: Use $f(t)=-k_b(x(t)-x_0)$ and  $S_t=\frac{v}{T}\int_0^tf(s)ds$.]}

{\bf c.} Demonstrate that $P(S_t)$ is a Gaussian distribution. Show that the Gaussian property of $P(S_t)$ implies Eq.\eqref{eq:fts:fdr}.{\it [Hint: To proof Gaussianity note that Eq.\eqref{eq:fts:lang} is linear with $x$.]}

\end{exercise}
We have remarked that Eq.\eqref{eq:fts:ft} is asymptotically valid in nonequilibrium steady states in the large time limit. Problem \ref{prob:FTs:steadystate} analyzes $P(S_t)$ for work $W_t$ and heat $Q_t$ in a steady state where the trap is moved at constant speed $v$. The main difference concerning the previous transient case is the initial condition. In the steady-state FT, the bead is already dragged by the moving trap at speed $v$. Instead, in the transient work-FT the bead is initially in equilibrium in a trap at rest, which is suddenly put in motion at the constant speed $v$. In the steady-state FT, $P(S_t=W_t/T)$ is a Gaussian at all times, however Eq.\eqref{eq:fts:ft} and Eq.\eqref{eq:fts:fdr} only hold in the limit $t/\tau_r \gg 1$. In contrast, in the transient work-FT $P(S_t=W_t/T)$ is also a Gaussian and Eq.\eqref{eq:fts:ftcrooks} and Eq.\eqref{eq:fts:fdr} are exact at all times. 

In general, for transient states (c.f. Eq.\eqref{eq:fts:ftcrooks}), Jarzynski's equality implies 
\begin{equation}
    \Delta G=-k_BT\log\langle \exp\bigl (-\frac{W}{k_BT}\bigr )\rangle
        \label{eq:fts:DG}
\end{equation}
showing that we can recover $\Delta G$ by exponentially averaging the irreversible work \cite{jarzynski2011equalities}. The average must be taken over an infinite number of experimental repetitions, the exponential average being strongly biased for a finite number of experiments \cite{palassini2011improving}. Equation \eqref{eq:fts:ftcrooks} permits us to extract less biased free energies of native structures from bi-directional pulling experiments, i.e., by combining unfolding (forward) and folding (reverse) work measurements \cite{collin2005verification}. In particular, Eq.\eqref{eq:fts:ftcrooks} predicts that $P_F(W)$ and $P_R(-W)$ cross at $W=\Delta G$. An extended version of Eq.\eqref{eq:fts:ftcrooks} \cite{junier2009recovery} has found important applications in molecular folding of nucleic acids \cite{alemany2012experimental,alemany2015free}, proteins \cite{alemany2016mechanical,rico2022molten} and ligand binding \cite{camunas2017experimental}. FTs have been also used to reconstruct folding free energy landscapes \cite{woodside2006direct}. They have also been used to infer free energy differences from partial work measurements, i.e., when full work cannot be measured, in yet another application of FTs to extract information about irreversible processes in small systems \cite{ribezzi2014free}.

\begin{figure}[b!]
	\begin{center}
		\includegraphics[height=4cm]{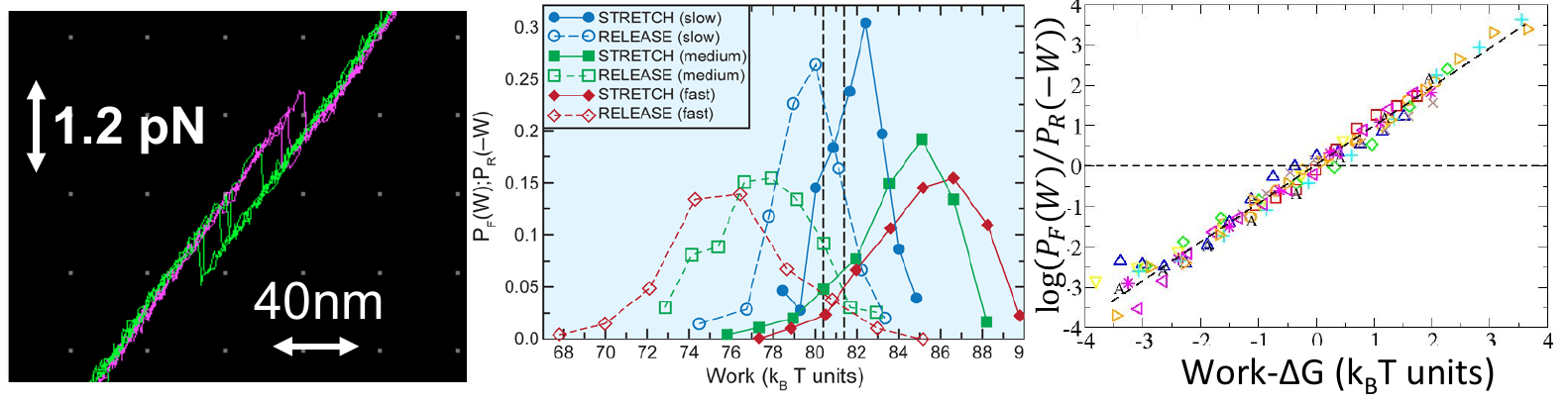}
	        \caption{
	        {\bf Free energy inference from nucleic acid hairpin pulling experiments.} (Left) Experimental force-distance curve of pulling a DNA hairpin with optical tweezers. The molecule unfolds around 16pN with unfolding (magenta) and refolding (green) trajectories exhibiting hysteresis. (Middle) Unfolding or forward (continuous lines) and refolding or reverse (dashed lines) work distributions at three different pulling rates. Hysteresis grows with the pulling speed. As predicted by Eq.\eqref{eq:fts:ftcrooks} unfolding and refolding work distributions cross at $\Delta G\sim 81k_BT$ for all pulling rates (1, 5, and 15 pN/s for blue, green, red, respectively). (Right) Experimental test of Eq.\eqref{eq:fts:ftcrooks}. Data for different molecules and speeds collapse into a straight line of slope $\sim 1$. Data from \cite{mossa2009dynamic}.  	    }\label{fig:FTs:Crooks}
	\end{center}
\end{figure}

The experimental measurement of energy fluctuations in molecular systems has opened new possibilities to explore living matter's marvelous performance. The successful exploitation of negative entropy production events during evolution has led to molecular structures of astonishing efficiency \cite{goldenfeld2011life}. By rectifying thermal fluctuations \cite{verley2014unlikely,schmitt2015molecular}, many enzymatic reactions such as molecular motors powered by ATP hydrolysis \cite{manosas2010active} and light-harvesting complexes in photosynthetic reactions can reach nearly 100$\%$ efficiencies. Such astonishing high efficiencies are not paralleled by man-built devices. The continuous remodeling of living matter by active weak molecular forces operating in the $k_BT$ scale, with their large intrinsic deviations and negative entropy production events, may have shaped biological matter over millions of years of evolution. Such questions touch the core of biology and lay open for future research.

\section{Energy, information and the Maxwell demon}
\label{FTs:energyinfo}
\begin{figure}[b!]
	\begin{center}
		\includegraphics[height=4cm]{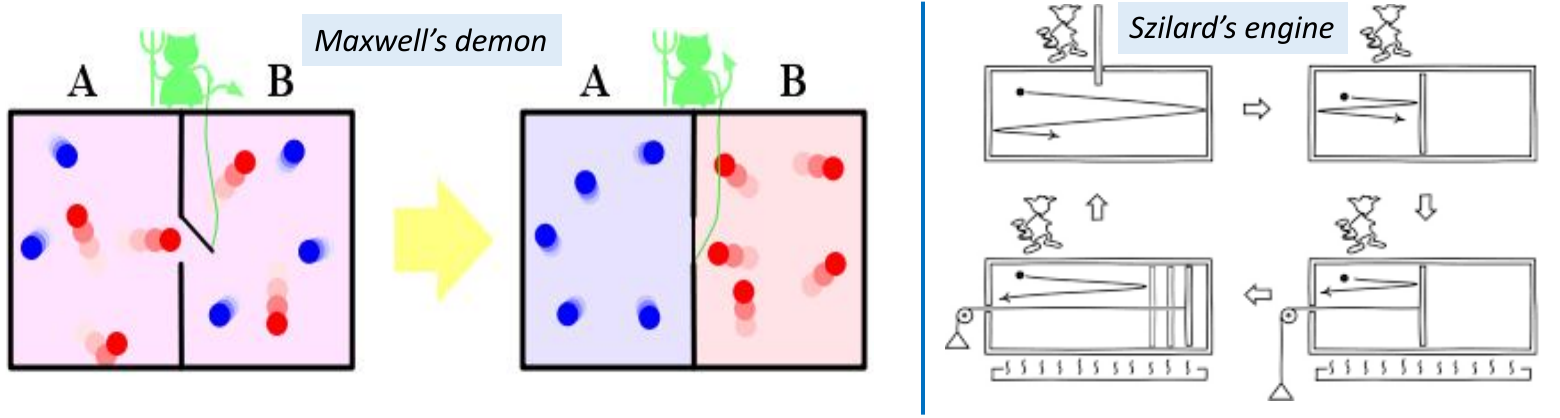}
	        \caption{
	        {\bf Information in thermodynamics.} (Left) Maxwell's demon schematics. Slow-moving molecules (blue) are separated from fast-moving ones (red) into the left and right compartments, creating a temperature difference without any work expenditure.  (Right) Szilard's engine schematics.  The trajectory of a single particle in two-compartment vessel of equal volumes is monitored. Work is extracted by inserting a wall in the middle and a pulley mechanism to convert heat from the bath into mechanical work implemented. A work equal to $k_BT\log 2$ is extracted by isothermically expanding the compartment containing the particle from half to full volume. An experimental realization of Szilard's engine is shown in Figure \ref{fig:FTs:DNASzilard}.}\label{fig:FTs:MDSZ}
	\end{center}
\end{figure}

In 1867 James Clerk Maxwell, the Scottish scientist who unified electricity and magnetism, proposed a thought experiment to violate the second law of thermodynamics \cite{maruyama2009colloquium}. Maxwell imagined a tiny intelligent being endowed with free will and fine enough tactile and perceptive organization to give him the faculty of observing and influencing individual molecules of matter \cite{bennett1987demons}. In his thought experiment (Figure \ref{fig:FTs:MDSZ}, left), two gas compartments are separated by an adiabatic wall and connected with a small hole and a gate the demon can open and close. By measuring the speed of the individual molecules, the demon selectively opens and closes the gate to separate the fast from the slow molecules creating a temperature gradient between the two compartments. The demon can do this effortlessly without the expenditure of work, violating the second law. 

\begin{figure}[b!]
	\begin{center}
		\includegraphics[height=9cm]{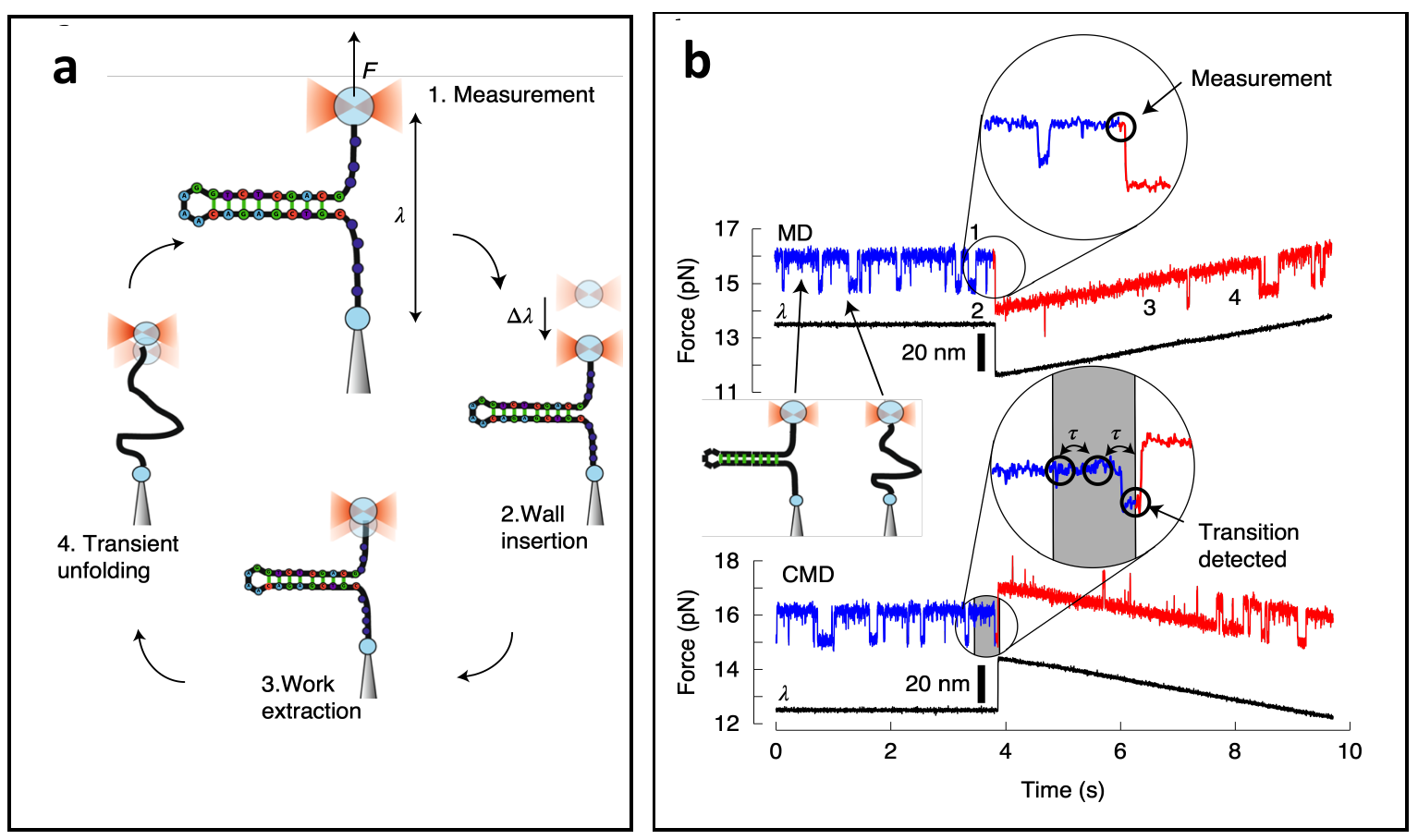}
	        \caption{
	        {\bf Experimental realization of Szilard's engine in the folding-unfolding of a DNA hairpin.} (Left) Schematics of the work extraction procedure. A DNA hairpin pulled by an optical trap hops between the folded and unfolded states. The molecular state is determined at a given time (step 1) and, depending on the outcome (folded or unfolded), the optical trap is instantaneously moved (folded, downwards; unfolded, upwards) a distance $\Delta\lambda$ to stabilize that state (step 2). By adiabatically re-positioning the original trap position (step 3), work is extracted from the transition events (step 4). In the figure, schematics are particularized for the case where the molecule is observed in the folded state. (Right) Force versus time trace for the single-measurement Szilard's engine (dubbed MD for Maxwell's demon, top) and for the multiple-measurements Szilard's engine (dubbed CMD for Continuous Maxwell's demon, bottom). In the CMD, measurements of the molecular state are made every $\tau$, and a work extraction protocol is implemented the first time the molecule is observed to change state (blue to red transition, circled zoomed regions). 	   	    }\label{fig:FTs:DNASzilard}
	\end{center}
\end{figure}

The solution to the paradox came in the '60s from the theory of computing developed by Landauer, Bennett and others \cite{leff2002maxwell}. They emphasized that recording information does not bring the system to its original state unless the information is erased. It is widely accepted that erasing bits produces heat that increases the entropy of the universe and restores the second law, however see \cite{ford2016maxwell}. Maxwell's demon can be experimentally realized in the Szilard engine (Figure \ref{fig:FTs:MDSZ}, right). A two-compartment box containing a single particle is in contact with a thermal bath at temperature $T$. An observation is made, and depending on which compartment the particle is in, a movable wall and a pulley mechanism that lifts a weight are implemented, fully converting heat from the bath into work. As before, the maximum extracted work per cycle equals the cost of erasing one bit (also called the Landauer limit), restoring the validity of the second law. The Szilard engine has been experimentally realized in small systems such as colloidal particles \cite{berut2012experimental,roldan2014universal}, electronic devices \cite{koski2014experimental}, and single molecules \cite{ribezzi2019large} (Figure \ref{fig:FTs:DNASzilard}). Crucial in the Szilard engine is the information content of the stored sequences ${\cal I}$ \cite{cover1999elements} which must be larger than the average extracted work according to the second law.

\begin{figure}[b!]
	\begin{center}
		\includegraphics[height=7cm]{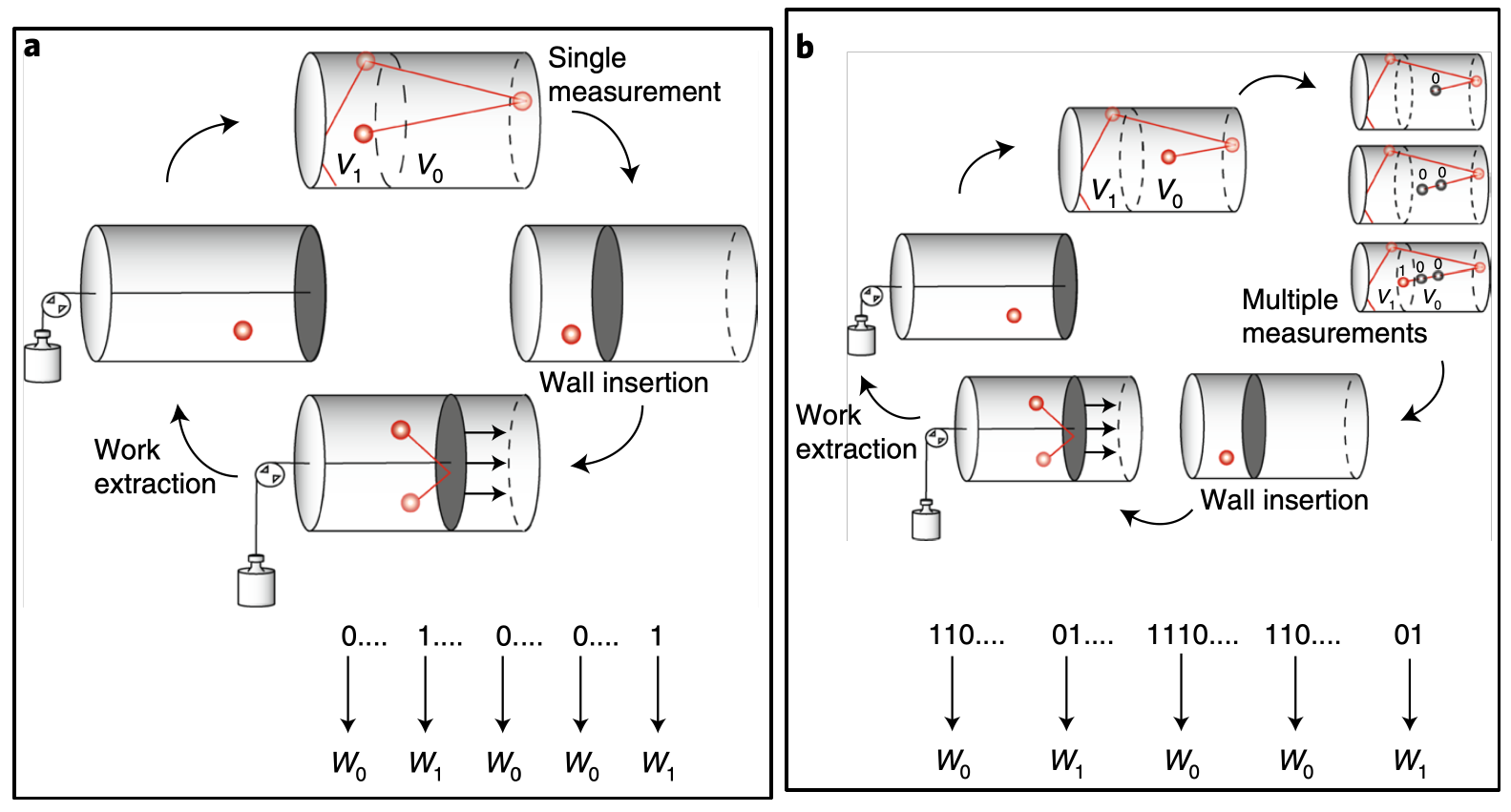}
	        \caption{
	        {\bf Single-measurement MD versus multiple-measurements CMD.} (Left) Stored sequences in the single-measurement MD consist of one bit ($\sigma=0,1$) depending on whether the particle is in the left or right compartment molecule, or folded and unfolded in Figure \ref{fig:FTs:DNASzilard}. The average work for a given $\sigma$-cycle equals $W_\sigma=-k_BT\log p_\sigma$ where $p_\sigma$ is the equilibrium probability of the measurement outcome $\sigma$ (see exercise \ref{ex:FTs:MD}. (Right) Stored sequences in the multiple-measurement CMD consist of strings of 2 or more bits such that all bits are identical and equal to $\sigma$ until the last one that changes to $1-\sigma$ (e.g. 0001,10,11110,00000001, etc.). 	    }\label{fig:FTs:MDCMD}
	\end{center}
\end{figure}

A continuous version of the Szilard engine, dubbed Continuous Maxwell demon (CMD), has been recently introduced where the particle's position is repeatedly monitored every time $\tau$ and work extracted only when the particle changes compartment \cite{ribezzi2019large,ribezzi2019work}. The average work per cycle has been shown to be larger than $k_BT\log 2$, see exercise \ref{ex:FTs:MD}. Because the stored sequences contain two or more bits, their average information content ${\cal I}$ is also larger than the Landauer limit $\log 2$, and the second law inequality, $W_{CMD}\le k_BTI$, preserved (see problem \ref{prob:FTs:MD}). The distinction between the single measurement MD and the multiple measurements CMD is shown in Figure \ref{fig:FTs:MDCMD},

\begin{figure}[b!]
	\begin{center}
		\includegraphics[height=6cm]{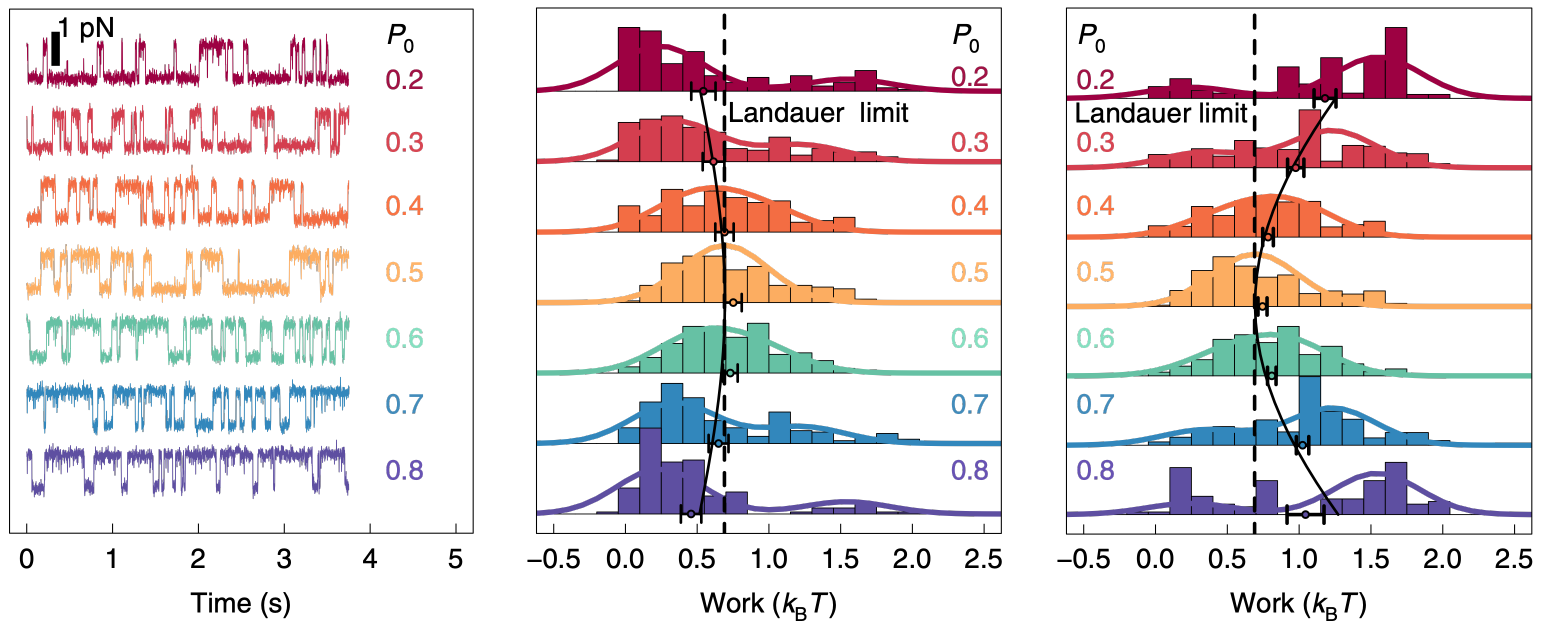}
	        \caption{
	        {\bf Work distributions in the MD and CMD for the single DNA hairpin unfolding experiments of Figure \ref{fig:FTs:DNASzilard} .} (Left) Experimental hopping traces at different $p_\sigma$ conditions. (Middle) Distributions of the extracted work per cycle for the single-measurement MD. Symbols are the average values, and the dashed line is the theoretical prediction, Eq.\eqref{eq:energyinfo:wmd}. (Right) Distributions of the extracted work per cycle for the multiple-measurement CMD. Symbols are the average values, and the dashed line is the theoretical prediction, Eq.\eqref{eq:energyinfo:wcmd}. 	    }\label{fig:FTs:workdistributions}
	\end{center}
\end{figure}

\begin{exercise}\label{ex:FTs:MD}
{\bf Information-to-energy conversion in the MD.} Let us consider the single-measurement (MD) and the multiple-measurement (CMD) depicted in Figure \ref{fig:FTs:MDCMD}. Do the following exercises,

{\bf a.} Show that the maximum average work that can be extracted when the particle is in compartment $\sigma=0,1$ (left or right) is given by $W_\sigma=-k_BT\log p_\sigma$, $p_\sigma$ being the probability to observe the particle in state $\sigma$ ($p_0+p_1=1$).

{\bf b.} Demonstrate that the maximum average work per cycle that can be extracted in the single-measurement MD is given by,

\begin{equation}
   W_{MD}=-k_B T\bigl(p_0\log p_0+p_1\log p_1 \bigr) \label{eq:energyinfo:wmd}
\end{equation}

whereas for the CMD with repeated measurements at every time $\tau$, 

\begin{eqnarray}
   W_{CMD}=-k_B T\bigl(p_0\log p_1+p_1\log p_0 \bigr) \label{eq:energyinfo:wcmd}
\end{eqnarray}

Explain why $W_{CMD}$ is independent of $\tau$.

{\bf c.} Demonstrate that $W_{MD}\le W_L\le W_{CMD}$ with $W_L=k_BT\log 2$ the Landauer limit. Show that, while $W_{MD}$ vanishes in the limit $p_0\to 0,1$, $W_{CMD}$ diverges in the same limit. Interpret this result.

{\bf d.} Figure \ref{fig:FTs:workdistributions} shows the experimentally measured work per cycle distributions in the MD and CMD. Can you explain the origin of the negative work events observed for CMD (right panel)? Note that a negative extracted work event means the extraction process is inefficient, i.e. heat is dissipated rather than used to extract work. 
\end{exercise}
The CMD has been experimentally implemented in single DNA hairpins mechanically unfolded and folded with optical tweezers (Figure \ref{fig:FTs:DNASzilard}), and the information content of the stored sequences analytically calculated for arbitrary $\tau$. In the limit of uncorrelated measurements $\tau\to\infty$, the average information content has been demonstrated to be minimum and larger than the extracted work, in agreement with the second law. Problem \ref{prob:FTs:MD} explains how to calculate the average information content per cycle of the stored sequences in the CMD in the limit $\tau\to\infty$, showing that it is larger than the average work per cycle, in agreement with the second law. A new class of generalized CMD models that combine Szilard and CMD-type protocols has been recently introduced, demonstrating that measurement correlations enhance information to energy conversion efficiency \cite{admon2018experimental,garrahan2023generalized}. The CMD has also been extended to N states \cite{raux2023n}.

The Szilard engine has endowed information of a thermodynamic meaning, extending fluctuation theorems to the case of feedback information \cite{sagawa2014thermodynamic,parrondo2015thermodynamics,schmitt2023information}. Feedback finds multiple applications in physics \cite{bechhoefer2005feedback,dieterich2016control} and quantum systems \cite{strasberg2022quantum}, it can be used to reduce dissipation \cite{tafoya2019using} and increase thermodynamic efficiency. Yet, we do not understand how feedback thermodynamics fits in a biological context. We distinguish discrete-time feedback (DTF) from continuous-time feedback (CTF). A system is driven out of equilibrium by the action of an external agent. In DTF, the driving protocol is changed depending on the outcome of one (or more) measurements taken at specific times \cite{sagawa2010generalized,toyabe2010experimental,potts2018detailed}. In CTF, an observable is continuously monitored, and the driving protocol is changed when the observable fulfills a specific condition \cite{horowitz2010nonequilibrium,rico2021dissipation}. For transient nonequilibrium states with feedback, the Jarzynski equality reads $\langle\exp(-W_{\rm d}/k_BT)\rangle=\gamma$, where $\gamma$ is called the efficacy and its logarithm $\Upsilon=\log\gamma$  thermodynamic information \cite{rico2021dissipation}. The convex property of the exponential function leads to $\langle W_{\rm d}\rangle\ge -k_BT\Upsilon$. For $\Upsilon\ge 0$ (information-to-work conversion), the dissipated work with feedback can even be negative but always larger than $-k_BT\Upsilon$ \cite{rico2021dissipation}. Protocols with $\Upsilon\ge 0$ are expected to reduce the overall dissipation of the nonequilibrium process, in contrast to those where $\Upsilon< 0$ incur additional losses and nothing is gained from feedback. Feedback protocols where dissipation is increased relative to the case without feedback might be called {\em unwise} protocols.  

\section{Outlook}
\label{FTs:infobio}
In 1944, Erwin Schroedinger published an enlightening monography titled ¨What is Life?¨ where he wrote \cite{schrodinger1944life}: The large and important and very much discussed question is: {\em How can the events in space and time which take place within the spatial boundary of a living organism be accounted for by physics and chemistry?} We accept that living beings do not violate fundamental laws of physics, yet they are of a particular kind. They seem to circumvent or mock the laws of physics as we understand them: a stone will fall if let fall due to gravity; however, birds fly whenever they decide so. Biologists have invented a term for this (teleonomy) to express the fundamental fact that living beings have their agenda, meaning that living beings move, eat, reproduce, play, do business, and so on \cite{kirschner2000molecular,pross2016life}. A key feature of nonequilibrium matter is entropy production equal to the rate at which heat is dissipated to the environment. Measuring entropy production rates has become a major challenge and a fundamental thermodynamic probe for life \cite{di2024variance,roldan2024thermodynamic}. Yet, we remain ignorant about which features of entropy production are intrinsic to life.   

Living matter is heterogeneous and soft \cite{kirkpatrick2015colloquium}, sharing features with active \cite{bechinger2016active} and disordered systems \cite{cocco2023statistical}. Cell populations are intrinsically heterogeneous: the same strain, environmental conditions, and “everything” often produce different outcomes. Often, there is no magic bullet to defeat dysfunctional masses of cells \cite{zhang2012physics}. At the molecular level, heterogeneity is widespread as many RNAs and proteins fold into a heterogeneous set of functional native structures \cite{solomatin2010multiple,liu2013dna}. Living matter is also soft, being actively remodeled by forces. These are strong enough to maintain stability but weak enough to allow for remodeling and adaptation. It is not accidental that the energy scale of thermal forces equals that of biochemical reactions ($1 k_BT=0.6$kcal/mol at 298K), reflecting that biological forces operate at the thermal noise level \cite{astumian2002brownian}. Most free energies of molecular structures are a few tens of kcal/mol or $k_BT$. As enthalpy ($\Delta H$) and entropy ($T\Delta S$) contributions are typically one order of magnitude higher than free energy ($\Delta G$) values, a compensation between enthalpy and entropy drives most molecular reactions (folding, binding, self-assembly, etc.).  Indeed, the enthalpy of a single hydrogen bond contributes with $\sim 7$kcal/mol, while the total enthalpy $\Delta H$ due to all bonds in an RNA or a protein can easily reach several hundreds of kcal/mol. Yet, the folding $\Delta G$ is typically ten or more times smaller due to a similar contribution by the entropy, $T\Delta S$.  This scenario describes protein folding in which, under appropriate conditions, the unfolded polypeptide chain collapses to the native state previous formation of a molten globule, a precursor of the native state in a funneled folding energy landscape \cite{bryngelson1995funnels,eaton1999searching,thirumalai2005rna,maity2005protein,rico2022molten}. Evolutionary forces have built biological systems over eons, from proteins to macromolecular complexes and beyond, driven by endogenous (internal) and exogenous (environmental) factors.  Operating over millions of years, these forces have modeled the things how we know them.  

It is soon to tell whether and how nonequilibrium physics will contribute to understanding the marvelous features of living matter \cite{ritort2022physics,bassereau2024physics}. The prominent role of information undeniably emerges from the experiments and theories developed over the past decades, a quantity that physicists identify with entropy but that calls for a broader ground to explain the complexity of biological matter \cite{mezard2009information}. One of the most appealing features of nonequilibrium work relations and fluctuation theorems is that they allow us to recover the second law as a particular case of more general mathematical equality \cite{jarzynski2011equalities}. This fact raises the intriguing question of whether the second law of thermodynamics is a disguised inequality of a yet unknown fundamental conservation law about energy and information \cite{ritort2016physics}.  In the best tradition of natural philosophy, experiments hand in hand with theory offer the best chances to unravel the secrets of life. The nonequilibrium physics of small systems is just a milestone in this exciting adventure.

\paragraph{Acknowledgements} I wish to thank the continued support of all {\em Small Biosystems Lab} members and collaborations at the University of Barcelona and abroad. I also express gratitude to the generous support of the Catalan Icrea Academia Institution through the Prizes 2008-2013-2018-2023 and the Spanish Research Council over the years, in particular grants PID2019-111148GB-100 and PID2022-139913NB-100.

\section*{Problems}\label{FTs:prob}


\begin{problem} 
\label{prob:FTs:WLC} 
{\bf Inextensible Worm-like-Chain.} In this problem, we reproduce the force-extension formula  Eq.\eqref{eq:sme:WLCMarkoSiggia} originally derived by John Marko and Eric Siggia in 1995 \cite{marko1995stretching}. The formula interpolates the low and high force regimes of the elastic rod model Eq.\eqref{eq:sme:WLCenergy} when a  force is applied to the polymer ends along the x-axis,
\begin{equation}
      \label{eq:sme:WLCenergyforce}
 E(\lbrace \hat{t}(s)\rbrace)=\frac{A}{2}\int_0^Lds\Bigl(\frac{\partial \hat{t}}{\partial s}\Bigr)^2 -f  \hat{x}\int_0^Lds\hat{t}(s) 
\end{equation}

The average extension is defined as,

\begin{equation}
 \label{eq:sme:extens}
    \langle x\rangle=-\frac{\partial \log {\cal Z}}{\partial f}=L\hat{x}\frac{\int [{\cal D}\hat{t}]\, \hat{t}(s)\exp(-\beta E(\lbrace \hat{t}(s)\rbrace))}{{\cal Z}}
\end{equation}

where $L$ is the contour length and ${\cal Z}=\int [{\cal D}\hat{t}]\exp(-\beta E(\lbrace \hat{t}(s)\rbrace))$ is the partition function with ${\cal D}\hat{t}$ denoting the path integral over $\hat{t}(s)$. To derive Eq.\eqref{eq:sme:WLCMarkoSiggia} we calculate the leading behavior of Eq.\eqref{eq:sme:extens} in the low and high $f$ regimes and interpolate between them. The low and high $f$ expansions are similar to the low and high-temperature expansions commonly employed in statistical mechanics (e.g., in the Ising model).  

{\bf a.} Expand Eq.\eqref{eq:sme:extens} around $f=0$ up to linear order in $f$ finding at low forces,

\begin{equation}
\label{eq:sme:extenslowf}
    \langle x\rangle=\frac{2LP}{3k_BT}f+{\cal O}(f^2)
\end{equation}

where $P=A/k_BT$ is the persistence length.  
{\it [Hint: Expand $\log {\cal Z}$ to first order in $f$ in Eq.\eqref{eq:sme:extens}. Evaluate the second cumulant of $E$ at $f=0$, $\langle E^2\rangle_{f=0}-\langle E\rangle^2_{f=0}$, using the zero force results of Exercise \ref{ex:FTs:WLC}.]}

{\bf b.} Expand Eq.\eqref{eq:sme:extens} around $\langle x\rangle=L$ and $f=\infty$ and find for the leading correction at high forces,

\begin{equation}
\label{eq:sme:extenshighf}
    \langle x\rangle=L\bigl [1-\sqrt{\frac{k_BT}{4Pf}}\bigr]+{\cal O}(\frac{1}{f})
\end{equation}  

{\it [Hint: This derivation is more elaborated. It requires decomposing $\hat{t}=\hat{t}_{\|}+\hat{t}_{\bot}$ in a longitudinal component, $\hat{t}_{\|}=(\hat{t}\cdot\hat{x})\hat{x}$, and a transverse component, $\hat{t}_{\bot}=(\hat{t}\cdot\hat{y})\hat{y}+(\hat{t}\cdot\hat{z})\hat{z}$. At high forces $|\hat{t}_{\bot}|\ll |\hat{t}_{\|}|$, we can express $\hat{t}$ as a function of the tranverse modes only, $\hat{t}=(1-|\hat{t}_{\bot}|^2/2)\hat{x}+\hat{t}_{\bot}$. By rewriting Eq.\eqref{eq:sme:WLCenergyforce} in terms of the Fourier transverse modes or discretizing the chain as suggested in Exercise \ref{ex:FTs:WLC} and using periodic Born-Von Karman conditions, we can integrate over the Gaussian transverse modes finding the above result.}

{\bf c.} From Eqs.\eqref{eq:sme:extenslowf},\eqref{eq:sme:extenshighf} derive the interpolation formula Eq.\eqref{eq:sme:WLCMarkoSiggia}. 

\end{problem}
\
\begin{problem} 
\label{prob:FTs:steadystate} 
{\bf Steady-State work fluctuation theorem.} Here, we will derive Eq.\eqref{eq:fts:ft} for the steady state of an optically trapped bead of friction coefficient $\gamma$ and dragged through water at speed $v$. For that we will consider Eq.\eqref{eq:fts:lang} and derive the steady-state (work) distribution for $W_t=\frac{v}{T}\int_0^tf(s)ds$ using $f(t)=-k_b(x(t)-x_0(t))$ with $x_0(t)=vt$ and $x(t)$ the fluctuating position of the bead.    

{\bf a.} Derive the autocorrelation function of the force in the steady state, $C(t)=\langle f(0)f(t)\rangle$, and compare the result with the equilibrium case, Eq.\eqref{eq:fts:cts}, noting that $\langle f(t)\rangle=\gamma v$ for all $t$. Show that the stationary distribution for the energy of the particle, $U(x)=(1/2)k_b(x-x_0)^2$ is given by the equilibrium Boltzmann distribution independently on the trap speed $v$, 

\begin{equation}
\label{eq:fts:PU}
    P(U)=\frac{1}{{\cal Z}}\exp\bigl(-\frac{U}{k_BT}\bigr)
\end{equation}

with ${\cal Z}=\sqrt{\frac{2\pi k_BT}{k_b}}$.

{\bf b.} Demonstrate that $P(W_t)$ is a Gaussian distribution that fulfills,

\begin{equation}
    \label{eq:fts:ssFT}
    \frac{P(W_t)}{P(-W_t)}=\exp\bigl( {\frac{W_t}{k_BT(1-\frac{\tau_r}{t}(1-e^{-\frac{t}{\tau_r}}))}}\bigr )
\end{equation}

Show that Eq.\eqref{eq:fts:fdr} holds in the limit $\frac{t}{\tau_r}\gg 1$ where the transient work-FT \eqref{eq:fts:ftcrooks} 
is recovered with $P_F=P_R$ and $\Delta G=0$.

{\bf c.} From the first law, $Q_t+W_t=\Delta U_t$, and Eqs.\eqref{eq:fts:PU},\eqref{eq:fts:ssFT} explain what should be the form of the heat distribution $P(Q_t)$. Show that $P(Q_t)$ has exponential tails. Do you think there is a corresponding FT for $Q_t$ in the large $t$ limit ($\frac{t}{\tau_r}\gg 1$)? {\it [Hint: The FT for heat is not as simple as with work. Keep your discussion at a qualitative level.]}

\end{problem}
\
\begin{problem} 
\label{prob:FTs:MD} 
{\bf Information-to-energy conversion in the Continuous Maxwell Demon (CMD).} In the CMD, a work extraction cycle consists of repeated measurements at regular time intervals $\tau$ until the particle changes compartment (Figure \ref{fig:FTs:MDCMD}, right) or the molecule changes state (Figure \ref{fig:FTs:DNASzilard}, bottom right). Stored sequences contain $n\ge 2$ bits, being of type ${\cal S}_n=\lbrace \sigma,\sigma,..,\sigma,1-\sigma \rbrace$ with $\sigma=0,1$. Let $p_{\sigma}$ be the probability of a measurement outcome, $\sigma$. In the limit of uncorrelated measurements ($\tau\to\infty$) the probability of a sequence ${\cal S}_n$ is given by, $P({\cal S}_n)=\prod_{k=1}^np_{\sigma_k}$. Do the following exercises,

{\bf a.} Calculate the average information content per cycle of the stored sequences,

\begin{equation}
   \label{eq:energyinfo:avinfo} 
   {\cal I}=-\sum_{n\ge 2}\sum_{{\cal S}_n}P({\cal S}_n)\log P({\cal S}_n)
\end{equation}

and demonstrate it is given by,

\begin{equation}
   \label{eq:energyinfo:avinfomin} 
   {\cal I}=-\sum_{\sigma=0,1}\bigl(\frac{p_{\sigma}}{p_{1-\sigma}}+p_{1-\sigma}\bigr)\log p_{\sigma}
\end{equation}

{\bf b.} Demonstrate the validity of the second law, 

\begin{equation}
    \label{eq:energyinfo:secondlaw} 
    W_{CMD}\le k_BT {\cal I}
\end{equation}

where $W_{CMD}$ and ${\cal I}$ are given in Eq.\eqref{eq:energyinfo:wcmd} and Eq.\eqref{eq:energyinfo:avinfomin} respectively.

{\bf c.} The information-to-energy thermodynamic cycle-efficiency is defined by,

\begin{equation}
    \label{eq:energyinfo:efficiency}
    \eta_I=\frac{W_{CMD}}{k_BT {\cal I}}
\end{equation}

Show that the minimum efficiency $\eta_I=1/3$ is obtained for $p_0=p_1=1/2$. Demonstrate that maximum efficiency $\eta_I=1$ is asymptotically obtained in the limit of rare events $p_0\to 0,1$. Show that in this limit both $W_{CMD}$ and  ${\cal I}$ diverge. Interpret this result. How does Eq.\eqref{eq:energyinfo:efficiency} compare with the efficiency of the standard MD, Eq.\eqref{eq:energyinfo:wmd}?
\end{problem}
%


\bibliography{biblio-contribution-FelixRitort}

\end{document}